\documentclass[11pt]{article}
\usepackage{amsfonts}
\def\lp{\stackrel{\leftarrow}{\partial}}

\def\be{\begin{eqnarray}}
\def\ee{\end{eqnarray}}
\def\*{\star}
\begin{document}
\begin{flushright}  %Nov 15, 2000 \\  hep-th/0011137 \\  
ANL-HEP-PR-00-115 \\ KUCP-171 \\  
Miami TH/2/00 
\end{flushright}

{\Large
\centerline{GENERATING ALL WIGNER FUNCTIONS}

Thomas Curtright$^{\S}$, Tsuneo Uematsu$^\natural$, and Cosmas Zachos$^{\P}$}\\ 

$^{\S}$ Department of Physics, University of Miami,
Box 248046, Coral Gables, Florida 33124, USA\\
\phantom{.} \qquad\qquad{\sl curtright@physics.miami.edu}  

$^{\natural}$ Department of Fundamental Sciences, 
FIHS, Kyoto University, Kyoto 606-8501, JAPAN\\ 
\phantom{.} \qquad\qquad{\sl uematsu@phys.h.kyoto-u.ac.jp} 

$^{\P}$ High Energy Physics Division,
Argonne National Laboratory, Argonne, IL 60439-4815, USA \\
\phantom{.} \qquad\qquad{\sl zachos@hep.anl.gov} 

%PACS: 
\begin{abstract} 
In the context of phase-space quantization, matrix elements and observables
result from integration of c-number functions over phase space, with Wigner
functions serving as the quasi-probability measure. The complete sets of
Wigner functions necessary to expand all phase-space functions include
off-diagonal Wigner functions, which may appear technically involved.
Nevertheless, it is shown here that suitable generating functions of these
complete sets can often be constructed, which are relatively simple, and lead
to compact evaluations of matrix elements. New features of such generating
functions are detailed and explored for integer-indexed sets, such as for the
harmonic oscillator, as well as continuously indexed ones, such as for the
linear potential and the Liouville potential. The utility of such generating
functions is illustrated in the computation of star functions, spectra, and
perturbation theory in phase space.
\end {abstract}
\noindent\rule{7in}{0.02in}
%\vskip 0.6cm

\section{Introduction}
General phase-space functions $f(x,p)$ and $g(x,p)$ compose noncommutatively
through Groenewold's $\star$-product \cite{groen}, which is the unique
associative pseudodifferential deformation \cite{bayen} of ordinary products:
\begin{equation}
\star \equiv ~e^{i\hbar(\stackrel{\leftarrow }{\partial }_{x}
\stackrel{\rightarrow }{\partial }_{p}-\stackrel{\leftarrow }{\partial }_{p}
\stackrel{\rightarrow }{\partial }_{x})/2}~.
\end{equation}
This product is the cornerstone of deformation (phase-space) quantization 
\cite{moyal,bayen,cfz,reviews}, as well as applications of matrix models and
non-commutative geometry ideas in M-physics \cite{natied}. Its mechanics,
however, is not always straightforward.

The practical Fourier representation of this product as an integral kernel has
been utilized widely since Baker's \cite{baker} early work,
\begin{equation} 
f\star g={1\over \hbar ^2 \pi^2}\int dp^{\prime}  dp^{\prime\prime}  dx' dx''  
~f(x',p')~g(x'',p'')~ \exp \left({-2i\over \hbar} 
\left( p(x'-x'') + p'(x''-x)+p''(x-x') \right )\right) .
\end{equation}
The determinantal nature of the star product controls the 
properties of the phase-space trace \cite{dbf,hansen},  
\begin{equation}
\int\! dp dx ~ f\star g = \int\! dp dx ~ fg
=\int\! dp dx ~ g\star f ~ \label{Ndjambi}.
\end{equation}

The above $\*$-product and phase-space integrals provide the multiplication law
and, respectively, the trace in phase-space quantization \cite{moyal}, the
third autonomous and logically complete formulation of quantum mechanics
beyond the conventional  formulations based on operators in Hilbert space or
path integrals. (This formulation is reviewed in \cite{bayen, reviews}.)
Properly ordered operators (e.g., Weyl-ordered) correspond uniquely to 
phase-space c-number functions (referred to as ``classical kernels" of the 
operators in question); operator products correspond to $\*$-products of 
their classical kernels;  and operator
matrix elements, conventionally consisting of traces thereof with the density
matrix, correspond to phase-space integrals of the classical kernels with the
Wigner function (WF), the Weyl correspondent of the density matrix
\cite{wigner,reviews}. The celebrated $\*$-genvalue functional equations
determining the Wigner functions \cite{dbf,dahl} and their spectral properties
(e.g.\ projective orthogonality \cite{takabayasi}) are reviewed and
illustrated in \cite{cfz}.

The functions introduced by Wigner \cite{wigner} and Szilard correspond to
diagonal elements of the density matrix, but quantum mechanical applications
(such as perturbation theory), as well as applications in noncommutative
soliton problems \cite{Gopakumar:2000zd}
often require the evaluation of off-diagonal matrix elements; they therefore
utilize the complete set of diagonal and off-diagonal generalized Wigner
functions introduced by Moyal \cite{moyal}. For instance, in noncommutative
soliton theory, the diagonal WFs are only complete for radial
phase-space functions (functions $\*$-commuting with the harmonic oscillator
hamiltonian---the radius squared), whereas deviations from radial symmetry
necessitate the complete off diagonal set.

As for any representation problem, the particular features of the $\*$-equations
under consideration frequently favor an optimal basis of WFs;
but, even in the case of the oscillator, the equations are technically
demanding. It is pointed out here, however, that suitable generating functions
for them, acting as a transform of these basis sets, often result in
substantially simpler and more compact objects, which are much easier to use,
manipulate, and intuit. Below, after some elementary overview of the Weyl
correspondence formalism (Sec 2), we illustrate such functions for the
harmonic oscillator (Sec 3), which serves as the archetype of WF 
bases indexed discretely; it turns out that these generating functions amount
to the phase-space coherent states for WFs, and also the WFs of coherent state
wavefunctions (Appendix A).  Direct
applications to first order perturbation theory are illustrated in Appendix B.

For sets indexed continuously, the generating function may range from a mere
Fourier transform, illustrated by the linear potential (Sec 4), to a 
less trivial continuous transform we provide for the Liouville potential 
problem (Sec 5), where the advantage of the transform method comes to cogent 
evidence. 

Throughout our discussion, we provide the typical $\*$-composition laws of such 
generating functions, as well as applications 
such as the evaluation of $\*$-exponentials of phase-space functions
 (Appendix C), or $\*$-versions of modified Bessel functions (techical 
aspects of integral transforms of which are detailed in Appendix D). 
Appendix E provides the operator (Weyl-) correspondent to the generating 
function for the Liouville diagonal WF introduced in Sec 5. 

\section{Overview of General Relations in the Weyl representation}

Without loss of generality, we review basic concepts in two-dimensional
phase space, $(x,p)$, as the extension to higher dimensions is straightforward.
In addition, we first address discrete spectra, $E_n$, $n=0,1,2,3,...$, 
and will only later generalize to continuous spectra.  

In the Weyl correspondence \cite{weyl}, c-number phase-space kernels $a(x,p)$
of suitably ordered operators ${\cal A(X,P)}$ are defined by 
\begin{equation}  
a(x,p)\equiv \frac{1}{2\pi} \int dy~ e^{-iyp} \langle x-\frac{\hbar}{2}y 
| {\cal A(X,P)} | x+\frac{\hbar}{2}y \rangle .
\end{equation}
Conversely, the ordering of these operators is specified through
\begin{equation}      
{\cal A}({\cal X},{\cal P}) 
=\frac{1}{(2\pi)^2}\int d\tau d\sigma 
dx dp ~a(x,p) \exp (i\tau ({\cal P}-p)+i\sigma ({\cal X}-x)). \label{corresp}
\end{equation}
An operator product then corresponds to a star-composition of 
these kernels \cite{groen},
\begin{equation}      
a(x,p)\* b(x,p)= \frac{1}{2\pi} \int dy~ e^{-iyp} \langle x-\frac{\hbar}{2}y 
| {\cal A(X,P) B(X,P)} | x+\frac{\hbar}{2}y \rangle ,
\end{equation}

Moyal \cite{moyal} appreciated that the density matrix in this 
phase-space representation is a hermitean generalization of the Wigner function:
\be
f_{mn}(x,p)&\equiv &\frac{1}{2\pi}\int dy~ e^{-iyp} \langle x-\frac{\hbar}{2}y 
| \psi_n \rangle \langle \psi_m | x+\frac{\hbar}{2}y \rangle  \nonumber\\
&=& \frac{1}{2\pi}\int dy e^{-iyp} \psi^*_m(x-\frac{\hbar}{2}y)
\psi_n(x+\frac{\hbar}{2}y)=f_{nm}^* (x,p)~,   \label{offdiag}
\ee
where the $\psi_m (x)$s are (ortho-)normalized solutions of a 
Schr\"{o}dinger problem. (Wigner \cite{wigner} mainly considered the diagonal 
elements of the density matrix 
(pure states), usually denoted as $f_m\equiv f_{mm}$.)
As a consequence, matrix elements of operators are produced by mere 
phase space integrals \cite{moyal},
\be
\langle \psi_m | {\cal A} |\psi_n \rangle =
\int dx dp ~a(x,p) f_{mn}(x,p).   \label{mamiwata}
\ee

The standard machinery of density matrices then is readily transcribed in
this language, e.g.\ the trace relation \cite{moyal},
\begin{equation} 
\int dxdp  ~f_{mn}(x,p)=
\int dx ~\psi^* _n(x) \psi_m (x) =\delta_{mn}~ ;     \label{trace}
\end{equation}
and \cite{dbf}
\begin{equation}                
f_{mn}\* f_{kl}=\frac{1}{2\pi\hbar}\delta_{ml}f_{kn} 
=\frac{1}{h}\delta_{ml}f_{kn} ~. \label{genstar}
\end{equation}
Given (\ref{Ndjambi}), it follows from the above that \cite{moyal}  
\begin{equation}                
\int dxdp ~f_{mn}(x,p)f^*_{lk}(x,p)=\frac{1}{2\pi\hbar}
\delta_{ml}\delta_{nk} ~.\label{ortho}
\end{equation}

For complete sets of input wavefunctions, it also follows that \cite{moyal}:
\be
\sum_{m,n} f_{mn}(x,p)f^*_{mn}(x',p')=\frac{1}{2\pi\hbar}
\delta(x-x')\delta(p-p')  ~.  \label{completeness}
\ee
An arbitrary phase-space function $\varphi(x,p)$ can thus be expanded as
\be
\varphi(x,p)=\sum_{m,n} c_{mn}f_{mn}(x,p),
\ee
the coefficients being specified through (\ref{ortho}),
\be
c_{mn}={2\pi\hbar}\int   dxdp\  f^*_{mn}(x,p)\varphi(x,p).
\ee
Further note the resolution of the identity \cite{moyal},
\begin{equation}    
\sum_nf_{nn}(x,p)=\frac{1}{2\pi\hbar}=\frac{1}{h} .   \label{resolution}
\end{equation}

For instance, for eigenfunctions of the hamiltonian ${\cal H(X,P)}$
with eigenvalues $E_n$, the corresponding WFs satisfy 
the following star-genvalue equations \cite{dbf} (further cf 
\cite{dahl,cfz}), with $H(x,p)$, the phase-space kernel of ${\cal H(X,P)}$:
\begin{equation}                
H\* f_{mn}=E_n f_{mn}, \qquad \qquad f_{mn}\* H=E_m f_{mn}.  \label{twosided}
\end{equation}

The time dependence of a pure state WF is given by Moyal's dynamical equation
 \cite{moyal}: 
\begin{equation} 
i\hbar \frac{\partial }{\partial t}\,f(x,p;t)=H\star f(x,p;t)-f(x,p;t)\star H.
\label{jomo}
\end{equation}
By virtue of the $\star$-unitary evolution operator (a ``$\star$-exponential"
\cite{bayen}),
\begin{equation} 
U_{\star} (x,p;t)=e_\star ^{itH/\hbar} \equiv 
1+(it/\hbar)H(x,p) + {(it/\hbar)^2\over 2!}  H\star H +{(it/\hbar )^3\over 3!} 
 H\star H\star H +...       ~,
\end{equation}
the time-evolved WF is obtained formally in 
terms of the WF at $t=0$, 
\begin{equation} 
f(x,p;t)=U_{\star}^{-1} (x,p;t) \star f(x,p;0)\star U_{\star}(x,p;t) .    
\label{legba}
\end{equation}
 (These associative combinatoric operations completely parallel those of 
operators in the conventional formulation of quantum mechanics in Hilbert 
space \cite{base}.)
Just like any star-function of $H$, this $\star$-exponential can be 
computed, \cite{bondia}
\begin{equation}                
\exp_\star(itH/\hbar)= \exp_\star(itH/\hbar) \* 1 
=\exp_\star(itH/\hbar)\* 2\pi\hbar \sum_n f_{nn} 
=2\pi\hbar \sum_n e^{it E_n/\hbar }f_{nn}.   \label{spectral}
\end{equation}
(Of course, for $t=0$, the obvious identity resolution is recovered).

For continuous spectra, the sums in the above relations extend to 
integrals over a continuous parameter (the energy), and the 
Kronecker $\delta_{mn}$s into $\delta$-functions (these last ones reflecting 
the infinite normalizations of unnormalizable states). E.g., 
eqns (\ref{trace}, \ref{ortho}) extend to
\be
\int dxdp ~f_{E_1E_2}(x,p)=\delta(E_1-E_2)\label{self-ortho},
\ee
\be
\int dxdp ~ f_{E_1E_2}(x,p)f^*_{E_1'E_2'}(x,p)=\frac{1}{2\pi\hbar}
\delta(E_1-E_1')\delta(E_2-E_2'),
\ee
Completeness (\ref{completeness}) extends to 
\be
\int dE_1 dE_2 ~f_{E_1E_2}
(x,p)f^*_{E_1E_2}(x',p')=\frac{1}{2\pi\hbar}
\delta(x-x')\delta(p-p').
\ee
More generally, (\ref{genstar}) extends to 
\be
f_{E_1E_2}\* f_{E_1'E_2'}=\frac{1}{2\pi\hbar}\delta(E_1-E_2')~ f_{E_1'E_2}.
\label{contgenstar}
\ee 
Finally, eqn (\ref{resolution}) extends to 
\be 
\frac{1}{2\pi \hbar}  &  = &
\frac{1}{2\pi}\int dy\,    e^{-ipy}\int dE\,
\langle x-\frac{\hbar y} {2} | E \rangle  \langle E |  x+\frac{\hbar y}{2} 
\rangle \nonumber \\  
&  = & \int dE ~ f_{EE} (x,p)  , 
\ee 
and hence (\ref{spectral}) extends to 
\begin{equation} 
\exp_\star \left( it H/ \hbar \right ) = 2\pi \hbar \int dE\, 
e^{itE/\hbar} \, f_{EE}\left(  x,p\right) \label{ctrace} . \end{equation}

\section{ Generating Functions for the Harmonic Oscillator}

Consider the harmonic oscillator,
\begin{equation}    
H(x,p)= \frac{1}{2} (p^2 + x^2),
\end{equation}
where, without loss of generality, parameters have been absorbed
in the phase space variables: $m=1$, $\omega=1$. Further recall that the 
normalized eigenfunctions of the corresponding operator hamiltonian 
${\cal H}$ are
$\psi_n(x)=\left({\sqrt{\pi} 2^nn!}\right)^{-\frac{1}{2}}
e^{-\frac{1}{2}x^2}H_n(x)$, for the eigenvalues $E_n= \hbar (n+1/2)$.  
Define a radial and an angular variable,
\begin{equation}               
z\equiv 4H=2(x^2+p^2) ~,   \qquad   \qquad   \tan\theta=\frac{p}{x} ~,
\end{equation}
so that 
\begin{equation}    
a {\sqrt{2}}\equiv 
(x+ip)=|x+ip|e^{i\theta}=\left(\frac{z}{2}\right)^{\frac{1}{2}}e^{i\theta}.
\end{equation}

Groenewold \cite{groen}, as well as Bartlett and Moyal \cite{bartlett}, 
have worked out the complete sets of solutions to Moyal's 
time-evolution equation (\ref{jomo}), which are all linear combinations 
of terms $\exp ({ it(m-n)})~ f_{mn}$. 
They solved that equation indirectly, 
by evaluating the integrals (\ref{offdiag}) for time-dependent Hermite 
wavefunctions, which yield generalized Laguerre 
polynomial-based functions. More directly, Fairlie \cite{dbf} dramatically 
simplified the derivation of the solution by relying on his fundamental 
equation (\ref{twosided}). He thus confirmed Groenewold's WFs 
\cite{groen,bartlett}, 
\be
f_{mn}(x,p)=\frac{(-1)^m}{\pi}\sqrt{\frac{m!}{n!}}
z^{\frac{n-m}{2}}e^{- z/2}
e^{i(n-m)\theta}L_m^{n-m}(z). \label{sesrumnir}
\ee

The special case of diagonal elements,
\begin{equation}
f_n\equiv f_{nn} 
=\frac{(-1)^n}{\pi}e^{-z/2}L_n(z),  
\end{equation}
constitutes the time-independent ``$\star$-genfunctions" of the oscillator
hamiltonian kernel \cite{cfz} (i.e.\, the complete set of solutions of
the time-independent Moyal equation $H\* f-f\* H=0$, where $H\* f_n= E_n f_n$.
 Incidentally, (\ref{genstar}) restricted to diagonal WFs 
closes them under $\*$-multiplication \cite{takabayasi}, 
$f_{m}\* f_{n}=\delta_{mn}f_{m} /(  {2\pi\hbar})$.)
That is to say, ``radially symmetric" phase-space functions,
i.e.\ functions that only depend on $z$ but not $\theta$, can be 
expanded in terms of merely these diagonal elements---unlike the most general 
functions in phase space which require the entire set of off-diagonal 
$f_{mn}$ above for a complete basis. Note, however, that all 
$\*$-products of such radially symmetric functions are commutative, 
since, manifestly, 
\begin{equation}                
\sum_n c_n f_n    \* \sum_m d_m f_m =\sum_m d_m f_m \* \sum_n c_n f_n  .
\end{equation}

Moreover, the $\*$-exponential (\ref{spectral}) for this set of 
$\*$-genfunctions is directly seen to amount to
\begin{equation}                
\exp_\star(itH/\hbar)= \left ( \cos(\frac{t}{2})\right ) ^{-1}
\exp\left ( \frac{2i}{\hbar} H\tan(\frac{t}{2})\right ) ,
\end{equation}
which is, to say, a Gaussian in phase space \cite{bayen}. As an application,
note that the hyperbolic tangent $\*$-composition law of gaussians follows 
trivially (since these amount to $\*$-exponentials with additive time 
intervals, $\exp_\star (tf)\* \exp_\star (Tf)=\exp_\star((t+T)f)$, \cite{bayen},
\begin{equation} 
\exp \left (-{a \over \hbar} (x^2+p^2)\right ) ~ \star ~ 
\exp \left (-{b \over \hbar} (x^2+p^2)\right ) = {1\over 1+ab} 
\exp \left (-{a+b\over\hbar (1+ab)} (x^2+p^2)\right ) .
\end{equation}

We now introduce the following generating function for the 
entire set of generalized Wigner functions,
\be
G(\alpha,\beta;x,p)& \equiv &\sum_{m,n}  \frac{\alpha ^m }{\sqrt{m!}}
\frac{\beta ^n}{\sqrt{n!}}f_{mn}
\nonumber\\
&&=\frac{1}{\pi}\sum_n \beta^n \frac{1}{n!}z^{n/2}e^{-z/2}
e^{in\theta}\sum_m \left( -z^{-\frac{1}{2}}e^{-i\theta} \alpha 
\right) ^mL_m^{n-m}(z) ~.
\ee
Utilizing the identity \cite{gradshteyn} 8.975.2, 
\be
\sum_{m=0}^\infty L_m^{n-m}(z) ~k^m=e^{-zk}(1+k)^n  ,
\ee
we obtain  
\be
G(\alpha,\beta;x,p)&=&\frac{1}{\pi}e^{-z/2}~
\sum_n \frac{1}{n!} \left(\beta \sqrt{z} e^{i\theta}\right)^n ~
e^{-z(-z^{-\frac{1}{2}}e^{-i\theta}\alpha)} ~
\left( 1-z^{-\frac{1}{2}} e^{-i\theta}\alpha \right ) ^n \nonumber\\
&=&\frac{1}{\pi}e^{-z/2}
\sum_n\frac{1}{n!}\left ( \beta \sqrt{z} e^{i\theta}-\alpha\beta\right ) ^n
e^{\sqrt{z} e^{-i\theta}\alpha}\nonumber\\
&=&\frac{1}{\pi}e^{-z/2}e^{\beta \sqrt{z} e^{i\theta}-\alpha \beta} ~
e^{\sqrt{z}  e^{-i\theta}\alpha}.
\ee
Thus, 
\be
G(\alpha,\beta;x,p)=\frac{1}{\pi}
\exp\left ( \sqrt{z} (\alpha e^{-i\theta}+\beta e^{i\theta})-\alpha \beta 
-\frac{z}{2}\right ).
\ee
Since 
\be
\sqrt{z} (\alpha e^{-i\theta}+\beta e^{i\theta})
=\sqrt{2}(\alpha+\beta )x-\sqrt{2}ip(\alpha-\beta)  ,
\ee
one can re-express: 
\be
G(\alpha ,\beta ;x,p) =G^{*}(\beta ,\alpha ;x,p)= \frac{1}{\pi}\exp\left (
\alpha \beta -\left(x-\frac{\alpha +\beta }{\sqrt{2}}\right)^2
-\left(p+i\frac{\alpha -\beta }
{\sqrt{2}}\right)^2
\right )  \label{gauss}.
\ee
As the name implies, from $G(\alpha,\beta;x,p)$, the $f_{mn}$s are 
generated by 
\be
f_{mn}(x,p)={1\over \sqrt{m! n!} }
\frac{\partial^m}{\partial \alpha^m}
\frac{\partial^n}{\partial \beta ^n} G(\alpha,\beta;x,p)\Biggl |_{\alpha=
\beta=0}.
\ee

These functions $\*$-compose as
\begin{equation}         
G(\alpha ,\beta )\* G(\epsilon ,\zeta )=
\frac{e^{\alpha\zeta}}{2\pi \hbar} ~ G(\epsilon ,\beta ).
\end{equation}
The phase-space trace is 
\begin{equation}         
\int dx dp ~ G(\alpha ,\beta ) = e^{\alpha \beta}.
\end{equation}
By (\ref{twosided}), the action of the Hamiltonian kernel on this 
function is  
\be
H\* G = \hbar\left ( {1\over 2}+\beta \frac {\partial}{\partial 
\beta }\right ) G = \hbar \left ({1\over 2} - \alpha \beta  
+ \beta  \sqrt{z} e^{i\theta} \right ) G,  \label{act1}
\ee
and
\be
G\* H = \hbar
\left (  {1\over 2} + \alpha \frac {\partial}{\partial \alpha } \right ) G 
= \hbar\left ({1\over 2}-\alpha \beta 
+ \alpha  \sqrt{z}  e^{-i\theta} \right ) G~.\label{act2}
\ee
Consequently,
\be
\int dx dp ~H\* G(\alpha,\beta)=   \hbar
\left (  {1\over 2} + \beta \frac {\partial}{\partial \beta }\right ) 
e^{\alpha \beta} =\hbar \left ( \frac{1}{2}+\alpha \beta \right ) 
e^{\alpha\beta}.
\ee
The spectrum then follows by operating on both sides of this equation,
\be
E_n&=& 
\frac{1}{n!}\frac{\partial^n}{\partial \alpha ^n}\frac{\partial^n}{\partial 
\beta ^n}
\int dx dp ~H\* G(\alpha ,\beta )\Biggl  |_{\alpha =\beta=0}\nonumber\\
&=& \frac{\hbar}{n!}\frac{\partial^n}{\partial \alpha ^n}
\frac{\partial^n}{\partial \beta ^n} \left(\frac{1}{2}+\alpha \beta \right)
e^{\alpha \beta }\Biggl |_{\alpha =\beta =0}=\hbar \left ( \frac{1}{2}+n
\right ) ~.
\ee
In general, matrix elements of operators may be summarized compactly through 
this generating function in phase space.

This generating function could be interpreted as a phase-space coherent state,
or the off-diagonal WF of coherent states, as 
discussed in Appendix A, \cite{dodonov}, 
\begin{equation}                
G(\alpha,\beta ;x,p)= \exp_\star(\beta a^\dagger)\ f_0\ \exp_\star( \alpha a ).
\end{equation}
\be
a\* G(\alpha,\beta)&=&\hbar \beta  G(\alpha ,\beta), \qquad 
a^\dagger\* G(\alpha ,\beta )=
\frac{\partial}{\partial \beta }G(\alpha ,\beta )\nonumber\\
G(\alpha ,\beta )\* a&=&\frac{\partial}{\partial \alpha }G(\alpha ,
\beta ), \qquad G(\alpha ,\beta )\* a^\dagger=\hbar \alpha  G(\alpha ,\beta ),
\ee
and hence eqs (\ref{act1},\ref{act2}) amount to 
\be
H\* G(\alpha ,\beta )&=&(a^\dagger\* a+\frac{\hbar }{2})\* 
G(\alpha ,\beta )=\hbar \left ( \beta \frac{\partial}{\partial \beta }
+\frac{1}{2} \right ) G(\alpha ,\beta ) \nonumber\\
G(\alpha ,\beta )\* H&=& G(\alpha ,\beta )\* (a^\dagger\* a+\frac{\hbar }{2})=
\hbar \left( \alpha \frac{\partial}{\partial \alpha }+\frac{1}{2}\right) 
G(\alpha ,\beta ).
\ee
This formalism finds application in, e.g., perturbation theory in
phase space, cf.\ Appendix B. 

\section{ Generating Functions for the Linear Potential}

The linear potential in phase space has been addressed \cite{dahl} (also 
see \cite{dodonov,cfz}).
We shall adopt the simplified conventions of \cite{cfz}, ie.\ 
$m=1/2$, $\hbar=1$. The Hamiltonian kernel is then,
\begin{equation}                
H(x,p)=p^{2}+x ~, 
\end{equation}
and the eigenfunctions of ${\cal H}$ are Airy functions,
\begin{equation}
\psi_{E}(x)=\frac{1}{2\pi }\int_{-\infty }^{+\infty}dX\,
e^{iX(E-x-X^2/3)}=\mathrm{Ai}(x-E)\;,
\end{equation}
indexed by the continuous energy $E$. The spectrum being continuous,
the Airy functions are not square integrable, but have continuum normalization,
$\int dx\,\psi_{E_{1}}^{\ast}\left( 
x\right)  \psi_{E_{2}}\left(  x\right)  =\delta\left(  E_{1}-E_{2}\right)$,
instead. Thus, (\ref{self-ortho}) et seq.\ are now operative.  
The generalized WFs are \cite{dahl}
\be
f_{E_1E_2}(x,p)&=&\frac{1}{4\pi^2}
\int dz e^{iz(\frac{E_1+E_2}{2}-x-p^2-z^2/12)}e^{ip(E_1-E_2)}
\nonumber\\
&=&e^{ip(E_1-E_2)}\frac{2^{2/3}}{2\pi}{\rm Ai}
\left(2^{2/3}(x+p^2-\frac{E_1+E_2}{2})\right).
\ee
The $\*$-exponential (\ref{ctrace}) then is again a plain exponential of the 
shifted hamiltonian kernel, 
\begin{equation}  
\exp_{\*} \left( it\left( x+p^{2}\right)
\right) =2\pi \int_{-\infty}^{\infty }dE\,e^{iEt}\,\frac{2^{2/3}}{2\pi
}\,\mathrm{Ai}\left( 2^{2/3}\left( x+p^{2}-\frac{E_{1}+E_{2}}{2}\right)
\right) =\exp \left( it\left( x+p^{2}+t^{2}/12\right) \right) . \label{linexp}
\end{equation}
(This could also be derived directly, as the CBH expansion simplifies 
dramatically in this case, cf.\ Appendix C.)
As before, the $\*$-composition law for plain exponentials of the hamiltonian 
kernel function follows,
\begin{equation}                
\exp\left(  a\left(  x+p^{2}\right)  \right)  \* \exp\left(  b\left(
x+p^{2}\right)  \right)  =\exp\left(  \left(  a+b\right)  \left( 
x+p^{2}-\frac{1}{4}ab\right)  \right). 
\end{equation}

Since the complete basis Wigner functions are now indexed continuously,
a generating function from them must rely on an integral instead of an
infinite sum. The simplest transform is possibly a double Fourier transform
with respect to the energy indices (but note the transform factors 
$\exp(i E_1 X),~ \exp(-iE_2 Y)$ may also be regarded as plane waves). 
Suitably normalized,
\be 
G\left( X,Y;x,p\right)& \equiv & 2\pi\int_{-\infty}^{+\infty}dE_{1} 
\int_{-\infty}^{+\infty}dE_{2}\;\left(  \frac{1}{\sqrt{2\pi}}e^{iE_{1} 
X}\right)  \,f_{E_{1}E_{2}}\left(  x,p\right)  \,\left( 
 \frac{1}{\sqrt{2\pi}}e^{-iE_{2}Y}\right)  \\   
& =& \frac{1}{2\pi}\int_{-\infty}^{+\infty}dE_{1}\int_{-\infty}^{+\infty} 
dE_{2}\;e^{i\left(  E_{1}-E_{2}\right)  \,p+iE_{1}X-iE_{2}Y}\;2^{2/3} 
\mathrm{Ai}\left(  2^{2/3}\left(  x+p^{2}-\frac{E_{1}+E_{2}}{2}\right) 
\right) \nonumber \\ 
&=& \frac{1}{2\pi}\int_{-\infty}^{+\infty}dE\int_{-\infty}^{+\infty} 
d\omega\;e^{i\omega\,p+i\left(  E+\omega/2\right)  X-i\left(  E-\omega  
/2\right)  Y}\;2^{2/3}\mathrm{Ai}\left(  2^{2/3}\left(  x+p^{2}-E\right)  
\right) \nonumber \\ 
&=&\delta\left(  p+\frac{X+Y}{2}\right)  \int_{-\infty}^{+\infty   
}dE\;e^{iE\left(  X-Y\right)  }\;2^{2/3}\mathrm{Ai}\left(  2^{2/3}\left( 
x+p^{2}-E\right)  \right)  \nonumber \\   
& =&\delta\left(  p+\frac{X+Y}{2}\right)  \int_{-\infty}^{+\infty   
}dE\;e^{iE\left(  X-Y\right)  }\;\frac{1}{2\pi}\int dz\,e^{iz(E-x-p^{2}  
-z^{2}/12)}  \nonumber  \\   
&=&\delta\left(  p+\frac{X+Y}{2}\right)  \;e^{i\left(  X-Y\right)  
(x+p^{2}+\left(  X-Y\right)  ^{2}/12)}. \nonumber     
\ee 

The phase-space trace is
\begin{equation} 
\int dx dp~ G(X,Y; x,p) =2 \pi \delta (X-Y)~,  \label{comp} 
\end{equation}
and, given (\ref{contgenstar}) for these functions, $
f_{E_1E_2}\* f_{E_1'E_2'}=$ $\frac{1}{2\pi}\delta(E_1-E_2')f_{E_1'E_2}$, 
 the $\*$-composition law for these $G$s is 
\begin{equation}     
G(X,Y; x,p) \* G(W,Z; x,p) 
= \delta (X-Z) ~G(W,Y;x,p).     \label{niflheimr}
\end{equation}

\section{ Generating Functions for the Liouville Potential}

A less trivial system with a continuous spectrum is the Hamiltonian 
with the Liouville potential, \cite{jackiw,Grosche}. In the conventions of 
\cite{cfz}, ($\hbar=1$, $m=1/2$), the Hamiltonian kernel is 
\begin{equation}                
H=p^2+e^{2x},
\end{equation}
and the eigenfunctions of the corresponding ${\cal H}$ are 
\begin{equation}                
\psi_E(x)=\psi_{E}^{*}(x)=
 \frac{1}{\pi}\sqrt{\sinh(\pi\sqrt{E})} ~K_{i\sqrt{E}}(e^x), \label{LioWaveFcn} 
\end{equation}
with continuum normalizations $\int dx\,\psi_{E_{1}}^{\ast}\left( 
x\right)  \psi_{E_{2}}\left(  x\right)  =\delta\left(  E_{1}-E_{2}\right)$.
The modified Bessel function (Cf.\ \cite{watson}, Ch VI, \S 6.22)
can be written in the Heine-Schl\"{a}fli form,
\be
K_{ip}( e^x) =\frac{1}{2}\int_{-\infty}^\infty  
dX\,\exp \left( -e^x \cosh X+iXp \right)
=K_{-ip}(e^{x})  . \label{watsonrep}
\ee

The non-diagonal WF is then
\be
f_{E_1E_2}(x,p)=
\frac{1}{\pi^3}\int dy e^{-2ipy}
\sqrt{\sinh(\pi\sqrt{E_1})}K_{i\sqrt{E_1}}^*(e^{x-y})
\sqrt{\sinh(\pi\sqrt{E_2})}K_{i\sqrt{E_2}}(e^{x+y}).
\ee
This Wigner function amounts to Meijer's $G$ function, 
\begin{equation}
f_{E_{1}E_{2}}\! (x,p)\!  =\!  
\frac{1}{8\pi^{3}}\sqrt{\sinh(\pi\sqrt{E_{1}})\sinh(\pi\sqrt{E_{2}})} 
\,G_{04}^{40}\left(  \frac{e^{4x}}{16}\left|  \frac{\,ip+\,i\sqrt{E_{1}}} 
{2},\frac{ip-\,i\sqrt{E_{1}}}{2},\frac{-ip+i\sqrt{E_{2}}}{2}\,,\frac 
{-ip-\,i\sqrt{E_{2}}}{2}\right.\right).
\end{equation} 

Alternatively, the WF may be written as a double integral representation,
\begin{eqnarray}
&&f_{E\left( k\right) \,E\left( q\right)}\left( x,p\right)= \label{alternate}\\ 
&&=\frac{1}{2\pi ^{3}}\sqrt{\sinh \left( \pi \sqrt{E\left( k\right) }\right) 
\sinh \left( \pi \sqrt{E\left( q\right) }\right) } 
\int\! dXdY \,e^{ikX}\,e^{iqY}\left( \frac{\cosh Y}{\cosh X}\right)^{ip}
K_{2ip}\left( e^{x}\sqrt{4\cosh X\cosh Y}\right) ,\nonumber   
\end{eqnarray}
where $E\left( k\right) \equiv k^{2},E\left( q\right) \equiv q^{2}$. 
This is an inverse integral transform, as in the preceding section, of a 
generating function 
\begin{eqnarray}
G\left( X,Y;x,p\right)  &\equiv &\int_{-\infty }^{\infty }\frac{dk}{\sqrt{
\sinh \left( \pi \sqrt{E\left( k\right) }\right) }}\int_{-\infty }^{\infty }
\frac{dq}{\sqrt{\sinh \left( \pi \sqrt{E\left( q\right) }\right) }}
~e^{-ikX}\,e^{-iqY}\,f_{E\left( k\right) \,E\left( q\right) }
\left( x,p\right)   \nonumber \\ 
&=&\frac{2}{\pi }\left( \frac{\cosh Y}{\cosh X}\right) ^{ip}\,K_{2ip}\left( 
e^{x}\sqrt{4\cosh X\cosh Y}\right) =G^{\ast }\left( Y,X;x,p\right) . 
\label{nondiagLio}
\end{eqnarray}
The form and construction of this $G$ are consequences of (\ref{watsonrep}), 
as detailed in Appendix D. 

However, the $\*$-composition law of this particular generating function is 
not so straightforward. It is singular, as a consequence of the 
general relation (\ref{contgenstar}) and the behavior of the integrand in 
(\ref{nondiagLio}) as $k,q\rightarrow 0$\footnote{ 
The singularity may be controlled by regulating the $\*$-product through 
imaginary shifts in the momenta,
$$
G\left( X,Y;x,p-\frac{i\epsilon }{2}\right) \*G\left( W,Z;x,p
+\frac{  i\epsilon }{2}\right) = 
\frac{1}{2\pi }G\left( W,Y;x,p\right) \Gamma \left( \epsilon \right)
\left ( e^{x}\sqrt{\frac{\cosh Y\cosh W}{\cosh X\cosh Z}}\left( \cosh 
X+\cosh Z\right) \right ) ^{-\epsilon }.
$$
It follows that one derivative with respect to either of $X$  
or $Z$ suffices to eliminate the divergence at $\epsilon =0$,
$$
\lim_{\epsilon \rightarrow 0}\partial _{X}G\left( X,Y;x,p-\frac{i\epsilon }
{2}\right) \*G\left( W,Z;x,p+\frac{i\epsilon }{2}\right)= 
$$
$$
=\frac{1}{2\pi }G\left( W,Y;x,p\right) 
\left( -\partial _{X}\right) \ln \left ( e^{x}\sqrt{\frac{\cosh 
Y\cosh W}{\cosh X\cosh Z}}\left( \cosh X+\cosh Z\right) \right ) 
=\frac{1}{2\pi }G\left( W,Y;x,p\right) \left( \frac{1}{2}\tanh X-\frac{
\sinh X}{\cosh X+\cosh Z}\right ).
$$
Unlike the situation in (\ref{niflheimr}), here the RHS vanishes at $X=Z$.
More symmetrically, 
$$
\lim_{\epsilon \rightarrow 0}\partial _{X}G\left( X,Y;x,p-\frac{i\epsilon 
}{2}\right) \*\partial _{W}G\left( W,Z;x,p+\frac{i\epsilon }{2}\right)
=\frac{1}{2\pi }\partial _{W}G\left( W,Y;x,p\right) \left\{ \frac{1}{2} 
\tanh X-\frac{\sinh X}{\cosh X+\cosh Z}\right\}.
$$
}.   %END OF FOOTNOTE

By some contrast to the above, an alternate generating function for 
just the diagonal WFs, $f_{EE}\equiv f_E$, could be defined through 
the spectral resolution of the $\* -K$ function, 
\begin{equation}
{\cal G}(z; x,p) \equiv K_{\* ~i\sqrt{H(x,p)}}\left(  e^{z}\right)
=2\pi\int_{0}^{\infty}  \! dE\, K_{i\sqrt{E}}  \left(  e^{z}\right) 
\,f_{E}(x,p)~.   \label{StarK} 
\end{equation}
This can be evaluated by reliance on Macdonald's trilinear 
identity \cite{watson,cataplex},
\begin{equation}                
\int_{0}^{\infty} \! dE ~K_{i\sqrt{E}} (e^{z})\;\psi_{E}(x)\,\psi_{E}^{\ast 
}(y)=\frac{1}{2}\,\exp\left(  -\frac{1}{2}\,\left(  e^{x+y-z}+e^{x-y+z}
+e^{-x+y+z}\right)  \right) .  \label{TrilinearMacdonald}   
\end{equation}
${\cal G} $ then is obtained by replacing 
$x\rightarrow x+Y$ and $y\rightarrow x-Y$, and Fourier 
transforming by $\frac{1}{\pi}\int \! dY ~e^{-2ipY}$, 
\begin{equation}                
\int_{0}^{\infty}dE\,K_{i\sqrt{E}} \left(  e^{z}\right) \,f_{E}\left( 
x,p\right)  
=\frac{1}{2\pi}\int dY\,e^{-2ipY}\, \exp\left(  -\frac{1}{2}\,
\left(  e^{2x-z}+e^{z+2Y}+e^{z-2Y}\right)  \right) .
\end{equation}
Finally, simplifying the RHS  gives
\be   
{2\pi} \int_{0}^{\infty} \! 
dE\,K_{i\sqrt{E}} \left(  e^{z}\right) \,f_{E}\left( x,p\right)  &  =& 
\exp\left(  -\frac{1}{2}\,e^{2x-z}\right)  \,\int \! dY\,e^{-2ipY} 
\,\exp\left(  -\frac{1}{2}\,e^{z}\left(  e^{2Y}+e^{-2Y}\right) \right)
\nonumber \\
&  =& \exp\left(  -\frac{1}{2}\,e^{2x-z}\right)  
\;K_{ip}\left( e^{z}\right) = {\cal G}(z; x,p) .   \label{DualTimeG} 
\ee 

As a side check of this expression, (\ref{DualTimeG}), note that it must 
satisfy the equations 
\be 
H\* \mathcal{G}(z;x,p)=\mathcal{G}(z;x,p)\* H=\left( -\partial  
_{z}^{2}+e^{2z}\right) \mathcal{G}(z;x,p)\;, \label{grendel}
\ee 
which follows from the spectral resolution evident in (\ref{StarK}). 
Indeed, since $e^{-z}\partial _{z}K_{ip}\left( e^{z}\right)   
=ipe^{-z}K_{ip}\left( e^{z}\right) -K_{ip+1}\left( e^{z}\right) $, and 
$\left( -\partial _{z}^{2}+e^{2z}\right) K_{ip}\left( e^{z}\right) 
=p^{2}K_{ip}\left( e^{z}\right) $, these relations are satisfied, 
\be 
\left( p^{2}+e^{2x}\right) \* \left( \exp \left( -\frac{1}{2} 
e^{2x-z}\right) K_{ip}\left( e^{z}\right)\right) &=&\left( \exp 
\left( -\frac{1}{2}
\,e^{2x-z}\right) K_{ip}\left( e^{z}\right)\right)  \* 
\left(p^{2}+e^{2x}\right) 
\nonumber \\  =\exp \left( -\frac{1}{2}\,e^{2x-z}\right) 
\left( -e^{2x-z}\,\partial_{z}K_{ip}\left( e^{z}\right) \right) 
&+&\left( p^{2}+\frac{1}{2}e^{2x-z}-\frac{1}{4}e^{4x-2z}\right) \exp \left( -
\frac{1}{2}\,e^{2x-z}\right) K_{ip}\left( e^{z}\right) \nonumber \\ 
&=&\left( -\partial _{z}^{2}+e^{2z}\right) \left( \exp \left( -\frac{1}{2}
\,e^{2x-z}\right) K_{ip}\left( e^{z}\right) \right) .
\ee

Parenthetically, as an alternative to the ordinary product form in 
(\ref{DualTimeG}), the phase-space kernel $\mathcal{G}$ may also be 
represented as an integral either of a $\*$-exponential or of a single 
$\*$-product\footnote{ NB Do not shift the integration parameter $y$ 
by the phase-space variable $x$ before the star products are evaluated. }, 
\begin{eqnarray}
\mathcal{G}(z;x,p) &=&\frac{1}{2}\int dy\,\exp _{\*}\left( -\frac{y}{2\sinh y}
e^{2x-z}+iyp-e^{z}\cosh y\right)  \nonumber \\ 
&=&\frac{1}{2}\int dy\,\exp \left( -\frac{1}{2}e^{y-z}e^{2x}\right) \*\exp 
(iyp-e^{z}\cosh y)\;.  \label{StarG} 
\end{eqnarray} 
This follows from the identities (cf.\ Appendix C), 
\be 
\exp _{\*}\left( -\frac{y}{2\sinh y}e^{2x-z}+iyp\right) =\exp 
\left( -\frac{1  }{2}e^{y-z}e^{2x}\right) \*\exp (iyp)
=\exp \left( -\frac{1}{2}    e^{2x-z}+iyp\right) \;.  \label{plechazunga} 
\ee 
The ordinary product form in (\ref{DualTimeG}) and the $\*$-exponential 
form in (\ref{StarG}) reveal that $\mathcal{G}(z;x,p)=\mathcal{G}(z;x,-p)$, 
so one may replace $\exp \left( iyp\right) $ by $\cos \left( yp\right) $ in
the second line of (\ref{StarG}) above. 
Given these, there are several ways to verify (\ref{grendel}). 
These relations and the star-product 
expressions for the kernel in (\ref{StarG}) are isomorphic to those of 
the corresponding operators, as discussed in Appendix E. 

The $\*$-composition law of these generating functions follows from 
(\ref{contgenstar}) and Macdonald's identity, 
\begin{equation}
\mathcal{G}(u;x,p)\* \mathcal{G}(v;x,p)={\frac{1}{2}}\int \!dw~\exp \left( 
-{ \frac{1}{2}}\left( e^{u+v-w}+e^{u-v+w}+e^{-u+v+w}\right) \right) ~ 
\mathcal{G }(w;x,p).  \label{calcomp} 
\end{equation} 
This also follows directly from the explicit form (\ref{DualTimeG}). Again, 
this is isomorphic to the corresponding operator composition law given in 
Appendix E. 

From the orthogonality of the $\psi _{E}$s, the diagonal WFs 
may be recovered by inverse transformation, 
\begin{equation}
f_{E}(x,p)=\int \!dz~\frac{\sinh (\pi \sqrt{E})}{2\pi ^{3}}~K_{i\sqrt{E} 
}(e^{z})~\mathcal{G}(z;x,p).
\end{equation}
This representation and the specific factorized $x,p$-dependence of 
$\mathcal{G}$ can be of considerable use, e.g., in systematically computing 
diagonal matrix elements in phase space.

In illustration of the general pattern, consider the first-order
energy shift effected by a perturbation Hamiltonian kernel $H_{1}$.
It is,  cf.\ Appendix B, eqn (\ref{energyshift}),  
\begin{equation}
\Delta E=\int \!dzdxdp~H_{1}~\frac{\sinh (\pi \sqrt{E})}{2\pi ^{3}}
K_{i\sqrt{E}}(e^{z})~\mathcal{G}(z;x,p).
\end{equation}
Choosing  
\begin{equation}                
H_{1}=e^{2nx}  e^{isp/2} ,
\end{equation}
yields 
\begin{equation} 
\Delta E
=\frac{\sinh (\pi \sqrt{E})}{2\pi ^{3}}~\int dz\,K_{i\sqrt{E} 
}(e^{z})\,\left( \int dx\,  e^{2nx}   \exp \left( -\frac{1}{2}
\,e^{2x-z}\right) \right) \,\left( \int dpK_{ip}\left( e^{z}\right)
\,   e^{isp/2} \right) .
\end{equation}

Now,
\begin{equation}
\int dx\,e^{2nx}\exp \left( -\frac{1}{2}\,e^{2x-z}\right) 
=2^{n-1}\Gamma \left(  n\right) \,e^{nz}~,
\end{equation}
and hence, (\cite{gradshteyn} 6.576.4, $a=b$),
\be 
&&  \int dz K_{i\sqrt{E}}(e^{z})\,K_{ip}\left( e^{z}\right) \,e^{nz} \\
&&  = \frac{
2^{n-3}}{\Gamma \left( n\right) }\,\Gamma \left( \frac{n+i\sqrt{E}+ip}{2}
\right) \Gamma \left( \frac{n+i\sqrt{E}-ip}{2}\right) \Gamma \left( 
\frac{n-i 
\sqrt{E}+ip}{2}\right) \Gamma \left( \frac{n-i\sqrt{E}-ip}{2}\right) 
.\nonumber 
\ee 
Thus, 
\be 
&& \Delta E=\frac{\sinh (\pi \sqrt{E})}{2\pi ^{3}}~4^{n-2} ~\times \\
&& \int  dp  e^{isp/2}\Gamma \left( \frac{n+i\sqrt{E}+ip}{2}\right) 
\Gamma \left( \frac{ n+i\sqrt{E}-ip}{2}\right) 
\Gamma \left( \frac{n-i\sqrt{E}+ip}{2}\right)
\Gamma \left( \frac{n-i\sqrt{E}-ip}{2}\right). \nonumber
\ee 
Finally, (\cite{gradshteyn} 6.422.19), 
\be 
\int dp\,e^{isp/2}\,\Gamma \left( \frac{n+i\sqrt{E}+ip}{2}\right) \Gamma 
\left( \frac{n+i\sqrt{E}-ip}{2}\right) \Gamma \left( \frac{n-i\sqrt{E}+ip}{2}
\right) \Gamma \left( \frac{n-i\sqrt{E}-ip}{2}\right) &&    \\ 
= 4\pi ~G_{22}^{22}\left( e^{s}\left| 
\begin{array}{c}
\frac{2-n\,+\,i\sqrt{E}}{2},\frac{2-n-\,i\sqrt{E}}{2} \\ 
\frac{n+i\sqrt{E}}{2}\,,\frac{n-\,i\sqrt{E}}{2}    
\end{array} 
\right. \right)  .&&  \nonumber
\ee 

To sum up, the perturbed energy shift is a Meijer function, 
\begin{equation}
\Delta E=\frac{4^{n} \sinh (\pi \sqrt{E})}{8\pi ^{2}}~ 
G_{22}^{22}\left( e^{s}\left| 
\begin{array}{c}
\frac{2-n\,+\,i\sqrt{E}}{2},\frac{2-n-\,i\sqrt{E}}{2} \\ 
\frac{n+i\sqrt{E}}{2}\,,\frac{n-\,i\sqrt{E}}{2}
\end{array}     
\right. \right) .
\end{equation}
In principle, any polynomial perturbation in either $x$ or $p$ can be
obtained from this, by differentiation with respect to $n$ and $s$. 
(Retaining  a bit of exponential in $x$ would be helpful to 
suppress the region of large negative $x$).

\vskip 1cm

\noindent{\Large{\bf Acknowledgments}}\newline 
We wish to thank D Fairlie and T Hakioglu for helpful conversations.
This work was supported in part by the US Department of Energy, 
Division of High Energy Physics, Contract W-31-109-ENG-38; NSF Award 0073390;
and by the Grant-in-Aid for Priority Area No 707 of the Japanese Ministry 
of Education. T U and T C thank Argonne National Laboratory for its hospitality 
in the summer of 2000.
\vskip 1cm

\appendix{\Large \bf Appendix A ~~~~$\*$-Fock Space and Coherent States}

Dirac's Hamiltonian factorization method for 
algebraic solution of the harmonic oscillator carries through 
(cf.~\cite{bayen})  intact in $\star$-space. Indeed,
\begin{equation}
H= \frac{1}{2} (x-ip) \star (x+ip) + \frac{\hbar }{2},
\end{equation}
motivating definition of 
\begin{equation}
a\equiv \frac{1}{\sqrt{2}} (x+ip),   \qquad \qquad 
a^{\dagger} \equiv \frac{1}{\sqrt{2}} (x-ip).
\end{equation}
Thus, noting
\begin{equation} 
a \star     a^{\dagger} -a^{\dagger} \star a=\hbar ,  
\end{equation}
and also that, by above, 
\begin{equation} 
a \star f_0= \frac{1}{\sqrt{2}} (x+ip) \star e^{-(x^2+p^2)}=0,
\end{equation}
provides a $\star$-Fock vacuum, it is evident that associativity 
of the $\star$-product permits the entire ladder spectrum 
generation to go through as usual. The $\star$-genstates of 
the Hamiltonian, s.t.~$H\star f= f\star H$, are thus
\begin{equation}
f_{nn}= f_n= \frac{1}{n!} (a^{\dagger}\star)^n   ~ f_0 ~(\star a)^n .
\end{equation}
These states are real, like the Gaussian ground state, and are thus 
left-right symmetric $\star$-genstates. They 
are also transparently $\star$-orthogonal for different eigenvalues; 
and they project to themselves, as they should, since the Gaussian ground 
state does, $f_0 \star f_0 =f_0/2\pi \hbar $. 

The complete set of generalized WFs can thus be written as 
\begin{equation}                
f_{mn}=\frac{1}{\sqrt{n! ~ m!}}(a^\dagger\*)^n 
f_0\   (\* a)^m \quad , \qquad   \qquad   m,n=0,1,2,3, \cdots
\end{equation}

The standard combinatoric features of conventional Fock space 
apply separately to left and right (its adjoint) $\*$-multiplication:
\be
&&a\* f_n\equiv a\* f_{nn}=\hbar \sqrt{n} f_{n,n-1}\nonumber\\
&&a^\dagger\* a\* f_n=\hbar \sqrt{n}a^\dagger\* f_{n,n-1}=\hbar nf_n\nonumber\\
&&a^\dagger\* f_n \equiv a^\dagger\* f_{nn}=\sqrt{n+1} f_{n,n+1}\nonumber\\
&&a\* a^\dagger\* f_n=\sqrt{n+1}~a\* f_{n,n+1}=\hbar (n+1)f_n   , 
\ee
\be
&&f_n\* a=\sqrt{n+1} f_{n+1,n}\nonumber\\
&&f_n\* a\* a^\dagger=\hbar (n+1)f_n\nonumber\\
&&f_n\* a^\dagger=\hbar \sqrt{n} f_{n-1,n}\nonumber\\
&&f_n\* a^\dagger\* a=\hbar nf_n   ~.  
\ee

Furthermore, a left/right (non-self-adjoint) coherent state is 
naturally defined \cite{dodonov,kim}
\be
\Phi(\alpha ,\beta )={\exp}_\star (\alpha a^\dagger)\ f_0\ 
{\exp}_\star (\beta a),  \qquad 
a\*\Phi(\alpha ,\beta )=\alpha \Phi(\alpha ,\beta ),\qquad 
\Phi(\alpha ,\beta )\* a^\dagger =\beta \Phi(\alpha ,\beta )\label{coher}.
\ee
Up to a factor of $\exp ((|\alpha|^2 + |\beta|^2)/2)$, this is also the 
WF of coherent states $|\alpha \rangle$ and $\langle \beta |$,
\cite{kim}. As indicated in the text, this coherent state is identifiable 
with the generating function $G$ for the harmonic oscillator. 

\appendix{\Large \bf Appendix B ~~~  Stationary Perturbation Theory} 

Perturbation theory could be carried out in Hilbert space and its resulting
wavefunctions utilized to evaluate the corresponding WF integrals.
However, in the spirit of logical autonomy of Moyal's formulation of Quantum
Mechanics in phase space, the perturbed Wigner functions may also be computed
ab initio in phase space \cite{bartlett,wangoconnell}, without reference to 
the conventional Hilbert space formulation. The basics are summarized below.
 
As usual, the Hamiltonian kernel decomposes into free and perturbed parts,
\be
H=H_0+\lambda H_1\label{hamil}~.
\ee
Fairlie's stationary, real, $\*$-genvalue equations \cite{dbf,cfz} 
for the full hamiltonian,
\be
H(x,p)\star f_n(x,p)=f_n(x,p)\star H(x,p)=E_n(\lambda) f_n(x,p),\label{stargen}
\ee
are solved upon expansion of their components $E$ and $f$ in powers of 
$\lambda$, the perturbation strength, 
\be
& E_n=&E^0_n+\lambda E^1_n+\lambda^2 E^2_n+ \cdots \\ 
&f_n=&f^0_n+\lambda f^1_n +\lambda^2 f^2_n+ \cdots .
\ee
Note the superscripts on $E$ and $f$ are order indices and not exponents. 
Resolution into individual powers of $\lambda$ yields the real equations: 
\be
&&H_0\* f^0_n= f^0_n\* H_0=E^0_nf^0_n  \label{L1}\\
&&H_0\* f^1_n+H_1\* f^0_n= f^1_n\* H_0+f^0_n\* H_1=
E^0_nf^1_n+E^1_nf^0_n\label{L2}\\
&&H_0\* f^2_n+H_1\* f^1_n=f^2_n\* H_0+f^1_n\* H_1=
E^0_nf^2_n+E^1_nf^1_n+E^2_nf^0_n ~.\label{L3}
\ee

Left multiplication of (\ref{L2}) by $f^0_n\*$ yields 
\be
f^0_n\* H_0\* f^1_n+f^0_n\* H_1\* f^0_n
=E^0_nf^0_n\* f^1_n+E^1_n f^0_n\* f^0_n ,
\ee
and, by (\ref{L1}),  
\be
f^0_n\* H_1\* f^0_n=E^1_n ~f^0_n\* f^0_n   ~;
\ee
by (\ref{ortho}, \ref{genstar}), and the cyclicity of the 
trace (\ref{Ndjambi}), 
\be
\int dx dp ~ E^1_n{f^0_n}\* f^0_n =\int dx dp (f^0_n\* H_1\* f^0_n)
=\int dx dp (H_1\* f^0_n\* f^0_n )= {1 \over 2\pi\hbar} 
\int dx dp H_1\* f^0_n ~.
\ee
Hence, 
\be
E^1_n=\int dx dp ~ H_1  f^0_n ,   \label{energyshift} 
\ee
the diagonal element of the perturbation. For the off-diagonal elements, 
similarly left-$\*$-multiply (\ref{L2}) by $f^0_m$, 
\be
f^0_m\* H_0\* f^1_n+f^0_m\* H_1\* f^0_n
=E^0_nf^0_m\* f^1_n+E^1_n f^0_m\* f^0_n   ~.
\ee

By completeness, $f^i_n$, $i \neq 0$, resolves to 
\be
f^i_n=\sum_{k,l} a^{i}_{n,kl}f^0_{kl}  ~, 
\ee
the reality condition dictating 
\be
a^i_{n,kl}=a^{*i}_{n,lk} ~.
\ee
Consequently, by 10, 
\be
E^0_m\sum_{k,l}a^1_{n,kl} f^0_m\* f^0_{kl}+f^0_m\* H_1\* f^0_n=
E^0_n\sum_{k,l}a^1_{n,kl} f^0_m\* f^0_{kl}+E^1_n\frac{1}{2\pi\hbar}
f^0_n\delta_{mn}~,
\ee
and hence 
\be
E^0_m\sum_{k}a^1_{n,km} f^0_{km}+2\pi\hbar f^0_m\* H_1\* f^0_n=
E^0_n\sum_{k}a^1_{n,km} f^0_{km}+E^1_n f^0_n\delta_{mn}~.
\ee

For $m\neq n$, 
\be
(E^0_n-E^0_m)\sum_{l}a^1_{n,lm} f^0_{lm}=2\pi\hbar (f^0_m\* H_1\* f^0_n),
\ee
so that 
\be
\sum_l a^1_{n,lm}f^0_{lm}=\frac{2\pi\hbar (f^0_m\* H_1\* f^0_n)}
{E^0_n-E^0_m} ~.
\ee
Finally, use of (\ref{ortho}), yields 
\be
a^1_{n,lm}&=&(2\pi\hbar)^2\int dxdp ~\frac{f^0_{ml}\*
f^0_m\* H_1\* f^0_n}{E_n^0-E_m^0}\nonumber\\
&=&(2\pi\hbar)^2\int  dxdp \frac{1}{2\pi\hbar} 
\frac{f^0_{ml}\* H_1\* f^0_n}{E_n^0-E_m^0}\nonumber\\
&=&(2\pi\hbar)\int dxdp 
\frac{H_1\* f^0_n\* f^0_{ml}}{E_n^0-E_m^0}\nonumber\\
&=&\frac {\delta_{nl}} {E_n^0-E_m^0} \int dxdp ~{H_1 f^0_{mn}} 
\qquad \qquad (m\neq n) ~.
\ee
We also have the similar equation for $l\neq n$.
Consequently, $a^{1}_{n,lm}$ is proportional to the matrix element of the 
perturbation, and it vanishes unless $l$ or $m$ is equal to $n$.
(NB This differs from \cite{wangoconnell} eqn (45).) To sum up,
\be
f^1_n= \sum_{m\neq n} \frac {1} {E_n^0-E_m^0}\left( 
 f^0_{nm} ( \int dx'dp' ~H_1(x',p') f^0_{mn}(x',p')) +
 f^0_{mn} (\int dx'dp' ~{H_1(x',p') f^0_{nm}(x',p')})  \right ).
\ee
By (\ref{mamiwata}), it can be seen that the same result may also follow
from evaluation of the WF integrals of perturbed wavefunctions
obtained in standard perturbation theory in Hilbert space.

For example, consider $H_1=\sqrt{2} ~x=a+a^\dagger$. It follows that 
$E^1_0=0$, and 
\be
a^1_{n,lm}&=&\frac{\delta_{n,l}}{(E^0_n-E^0_m)}\int\kern-0.6em\int dxdp
f^0_{mn}\* (a+a^\dagger)\nonumber\\
&=&\frac{\delta_{n,l}}{(E^0_n-E^0_m)}\int\kern-0.6em\int dxdp
(\sqrt{m+1}~f^0_{m+1,n}+\sqrt{n+1} ~f^0_{m,n+1})\nonumber\\
&=&\delta_{n,l}(\sqrt{m+1} ~\delta_{m+1,n}-\sqrt{n+1} ~\delta_{m,n+1})~,
\ee
for $m\neq n$ and the $(m\leftrightarrow l)$ expression for $l\neq n$. Hence, 
\be
f^1_n =\sqrt{n}(f^0_{n-1,n}+f^0_{n,n-1})-\sqrt{n+1}(f^0_{n,n+1}+f^0_{n+1,n}).
\ee

\appendix{\Large \bf Appendix C ~~~Combinatoric Derivation of Identities 
(\ref{linexp}) and (\ref{plechazunga}) }

The $\*$-exponential (\ref{linexp})  of the Hamiltonian kernel for the 
linear potential is also easy to work out directly, since the combinatorics 
in $\*$-space
are identical to the combinatorics of any associative algebra. In particular,
the Campbell-Baker-Hausdorff expansion also holds for $\*$-exponentials,
\begin{equation}                
\exp_{\* }\left(  A\right)  \* \exp_{\*}\left(  B\right)
=\exp_{\*}\left( 
A+B+\frac{1}{2}\left[  A,B\right]  _{\*}+\frac{1}{12}\left[  A,\left[ 
A,B\right]  _{\*}\right]  _{\*}+\frac{1}{12}\left[  \left[ 
A,B\right]  _{\*},B\right]  _{\*} + C \right),
\end{equation}
Where $C$ represents a sum of triple or more nested $\*$-commutators 
(Moyal Brackets, $[ A, B]_{\*} \equiv A\*B-B \* A$). Now, choosing  
$A=itx$ and $B=itp^{2}+it^{2}p+\frac{1}{3}it^{3}$,
yields $\left[  A,B\right]  _{\*}=-2it^{2}p -it^{3}$, $\left[ 
A,\left[  A,B\right]  _{\*}\right]  _{\*}=2it^{3}$, $\left[  
\left[  A,B\right]  _{\*} ,B\right]  _{\*}=0$, and hence $C=0$. 

Consequently,
\begin{equation}
\exp_{\*}\left( itx\right)  \*\exp_{\*}\left(  
itp^{2}+it^{2}p+\frac{1}{3}it^{3}\right)  
=\exp_{\*}\left(itx+itp^{2}\right)  \label{OpIdentity}. 
\end{equation}
But further note $\exp_{\*}\left(  ax\right) 
=\exp\left(  ax\right)$, and also $\exp_{\*}\left(  bp^{2}+cp+d\right) 
=\exp\left(  bp^{2}+cp+d\right)  $.   
This reduces the $\*$-product to a mere translation, 
\be 
\exp_{\*}\left(  ax\right)  \*\exp_{\*}\left( 
bp^{2}+cp+d\right)   &  =&\exp\left(  ax\right)  \*\exp\left(
bp^{2}+cp+d\right)  \\
&  =&\exp\left(  ax+\frac{1}{2}ia\partial_{p}\right)\exp\left(bp^{2}
+cp+d\right)\nonumber   \\
&=&\exp\left(  ax+b\left(  p+\frac{1}{2}ia\right) ^{2}+c\left(  p+\frac{1}{2}
ia\right)  +d\right)\nonumber   \\
&=&\exp\left(  ax+bp^{2}+\left(  c+iab\right)  p+d-\frac{1}{4}a^{2}b+\frac{1}
{2}iac\right). \nonumber 
\ee 
Consequently, 
\begin{equation} 
\exp_{\*}\left( itx\right)  \*\exp_{\*}\left( 
itp^{2}+it^{2}p+\frac{1}{3}it^{3}\right)  =\exp\left( it\left( 
x+p^{2}+t^{2}/12\right)  \right)  \label{StarExpsToOrdinary},
\end{equation}
and the identity 
$$
\qquad \qquad \qquad \exp_{\*}\left( it\left(  x+p^{2}\right)  \right)  
=\exp\left( it\left(  x+p^{2}+t^{2}/12\right)  \right)
\qquad \qquad \qquad \qquad \qquad \qquad \qquad (\ref{linexp})   
$$ 
follows. 

The proof of 
$$
\qquad \exp _{\*}\left( -\frac{y}{2\sinh y}e^{2x-z}+iyp\right) =\exp 
\left( -\frac{1  }{2}e^{y-z}e^{2x}\right) \*\exp (iyp)
=\exp \left( -\frac{1}{2}    e^{2x-z}+iyp\right) 
\qquad \qquad (\ref{plechazunga}) 
$$ 
is similar. Choosing now 
$A= -\frac{1  }{2}e^{y-z}e^{2x}$ and $B=iyp$, it follows that
$\left[  A,B \right] _{\*} =-2yA$, so that only those multiple Moyal 
commutators survive which are linear in $A$. This means, then, that in the 
Hausdorff expansion \cite{magnus} for 
$Z(A,B) \equiv \ln_\* \left( \exp_\* (A) \* \exp_\* (B)\right)$, only $B$
and terms {\em linear} in $A$ survive. Hence, $Z$ reduces to merely 
\be
Z= B + A \left (    \frac{B ] _{\*}} {1- e^{-B] _{\*}   }  }
\right ).
\ee 
The Hadamard expansion in the right parenthesis means successive 
right $\*$-commutation with respect to $B$ as 
many times as the regular power 
expansion of the function in the parenthesis dictates. Consequently, 
\be
\exp \left( -\frac{1  }{2}e^{y-z}e^{2x}\right) \*\exp (iyp)
=\exp_\* \left( -\frac{1  }{2}e^{y-z}e^{2x}\right) \*\exp_\* (iyp)=
\exp _{\*}\left( -\frac{y}{2\sinh y}e^{2x-z}+iyp\right). 
\ee
On the other hand, 
\be
\exp \left( -\frac{1  }{2}e^{y-z}e^{2x}\right) \*\exp (iyp)=
\exp \left( -\frac{1  }{2}e^{y-z +2x}\right)  
\exp \left (iy(p+i\lp_x /2)\right)
=\exp \left( -\frac{1}{2}    e^{2x-z}+iyp\right), 
\ee
and the identity is proven.

\appendix{\Large \textbf{Appendix D ~~~Construction of the 
Generating Function for the \phantom{yabadabadoo booo} Liouville WFs}}

From (\ref{LioWaveFcn}) and (\ref{watsonrep}), it is evident that the 
Liouville wave functions can be generated by 
\be  
\exp \left( -e^{x}\cosh X\right) =\int_{-\infty }^{\infty }\frac{dk}{\sqrt{
\sinh \left( \pi \sqrt{E\left( k\right) }\right) }}\,e^{-ikX}\,\psi
_{E\left( k\right) }(x),
\ee
where $E\left( k\right) \equiv k^{2}$. Therefore, the usual wave function 
bilinears appearing in the WFs are 
generated by (recalling that the  $\psi ^{\prime }s$ are real) 
\be 
&&\exp \left( -e^{x-y}\cosh X\right) 
\exp \left( -e^{x+y}\cosh Y\right)  \\
&&=\int_{-\infty }^{\infty }\frac{dk}{\sqrt{\sinh \left( \pi \sqrt{E\left( 
k\right) }\right) }}\int_{-\infty }^{\infty }\frac{dq}{\sqrt{\sinh \left( 
\pi \sqrt{E\left( q\right) }\right) }}\,e^{-ikX-iqY}\,\psi _{E\left(
k\right) }(x-y)\psi _{E\left( q\right) }(x+y). \nonumber 
\ee 
Consequently, Fourier transforming this produces a generating function 
for WFs,  
\be 
&&\frac{1}{\pi }\int_{-\infty }^{\infty }   dy ~e^{-2ipy}\exp \left( 
-e^{x-y}\cosh X\right) \exp \left( -e^{x+y}\cosh Y\right)  \\ 
&&=\int_{-\infty }^{\infty }\frac{dk}{\sqrt{\sinh \left( \pi \sqrt{E\left( 
k\right) }\right) }}\int_{-\infty }^{\infty }\frac{dq}{\sqrt{\sinh \left( 
\pi \sqrt{E\left( q\right) }\right) }}\,e^{-ikX-iqY}\,f_{E\left( k\right) 
\,E\left( q\right) }(x,p).   \nonumber 
\ee 
Evaluation of this expression yields just a factor multiplying 
a modified Bessel function,  
\be 
&&\int_{-\infty }^{\infty }dy ~e^{-2ipy}\exp \left( -e^{x-y}\cosh 
X-e^{x+y}\cosh Y\right)  \\ 
&&=\int_{-\infty }^{\infty }dy ~e^{-2ip\left( y+\frac{1}{2}\ln \left( \cosh
X/\cosh Y\right) \right) }\exp \left( -\left( e^{x}\sqrt{4\cosh X\cosh Y}
\right) \cosh y\right)\nonumber   \\
&&=2\left( \frac{\cosh Y}{\cosh X}\right) ^{ip}K_{2ip}\left( e^{x}\sqrt{
4\cosh X\cosh Y}\right). \nonumber 
\ee 

Thus, a generating function for the complete set of Liouville Wigner functions 
is 
\be 
&&\frac{2}{\pi }\left( \frac{\cosh Y}{\cosh X}\right) ^{ip}K_{2ip}\left(
e^{x}\sqrt{4\cosh X\cosh Y}\right)  \qquad \qquad \qquad \qquad \qquad 
\qquad \qquad \qquad (\ref{nondiagLio})
\nonumber \\
&&=\int_{-\infty }^{\infty }\frac{dk}{\sqrt{\sinh \left( \pi \sqrt{E\left(
k\right) }\right) }}\int_{-\infty }^{\infty }\frac{dq}{\sqrt{\sinh \left(
\pi \sqrt{E\left( q\right) }\right) }}\,e^{-ikX-iqY}\,f_{E\left( k\right)
\,E\left( q\right) }(x,p), \nonumber 
\ee 
as in the text.

\appendix{\Large \textbf{Appendix E ~~~Operator Ordering and eqn 
(\ref {DualTimeG})}}

Given the factorized phase-space generating function 
$$
\mathcal{G}(z;x,p)=\exp \left( -\frac{1}{2}\,e^{2x-z}\right) \;K_{ip}\left(
e^{z}\right),  \qquad \qquad \qquad (\ref {DualTimeG}) 
$$
what is the operator corresponding to it?
According to Weyl's prescription, eq (\ref{corresp}), the associated 
operator is 
\be 
\mathfrak{G}(z;\mathcal{X},\mathcal{P}) &=&\frac{1}{(2\pi )^{2}}\int d\tau
d\sigma dxdp~\mathcal{G}(z;x,p)\exp (i\tau (\mathcal{P}-p)+i\sigma 
(\mathcal{  X}-x)) \\   
&=&\frac{1}{(2\pi )^{2}}\int d\tau d\sigma dx dp~ 
\exp (i\tau \mathcal{P}+i\sigma \mathcal{X})\,
\exp \left( -\frac{1}{2}\,e^{2x-z}-i\sigma x\right)
K_{ip}\left( e^{z}\right)~ \exp (-i\tau p). \nonumber  
\ee 
The integrals over $x$ and $p$ may be evaluated separately, if the $\sigma $ 
contour is first shifted slightly above the real axis, 
$\sigma \rightarrow \sigma +i\epsilon $, thereby suppressing contributions to 
the $x$-integral as $x\rightarrow -\infty $. Now 
$s\equiv \frac{1}{2} e^{2x-z}$ gives
\be 
\int_{-\infty }^{+\infty }\!\! dx\exp \left( -\frac{1}{2}\,e^{2x-z}-i\left( 
\sigma +i\epsilon \right) x\right) =\int_{0}^{\infty }\frac{ds}{2s}\left( 
2se^{z}\right) ^{-i\left( \sigma +i\epsilon \right) /2}\exp \left( -s\right) 
&& \nonumber\\
=\frac{1}{2}\,e^{-i\left( z+\ln 2\right) \sigma /2}\,\Gamma \left( -i\left(
\sigma +i\epsilon \right) /2\right). &&
\ee 
By (\ref{watsonrep}), 
\be 
\int dpK_{ip}\left( e^{z}\right) \exp (-i\tau p)=\frac{1}{2}\int_{-\infty 
}^{\infty }dX\,e^{-e^{z}\cosh {X}}\,2\pi \delta \left( X-\tau \right) =\pi 
\,e^{-e^{z}\cosh {\tau }} .
\ee  
So 
\be 
\mathfrak{G}(z;\mathcal{X},\mathcal{P})=\frac{1}{8\pi }\int d\tau d\sigma
\,e^{-i\left( z+\ln 2\right) \sigma /2}\,\Gamma \left( -i\left( \sigma
+i\epsilon \right) /2\right) \,e^{-e^{z}\cosh {\tau }}\,\exp (i\tau \mathcal{
P}+i\sigma \mathcal{X}).
\ee
The shifted $\sigma $ contour avoids the pole in $\Gamma $ at the origin. 

Ordering with all $\mathcal{P}$s to the right, thereby departing from Weyl 
ordering, yields $\exp (i\tau \mathcal{P}+i\sigma \mathcal{X})=\exp (i\sigma 
\mathcal{X})\exp \left( i\sigma \tau /2\right) \exp (i\tau \mathcal{P})$. 
Performing the $\sigma $ integration before the $\tau $ integration, 
permits taking the limit $\epsilon \rightarrow 0$ to obtain  
\be  
\mathfrak{G}(z;\mathcal{X},\mathcal{P}) &=&\frac{1}{8\pi }\int\! d\tau 
\left( \int\! d\sigma \Gamma \left( -i\left( \sigma +i\epsilon \right) 
/2\right) \exp (i\sigma \mathcal{X}+i\sigma \tau /2-i\sigma \left( z+\ln 2
\right) /2)\right ) \,e^{-e^{z}\cosh {\tau }}\exp (i\tau \mathcal{P})
  \nonumber \\ 
&=&\frac{1}{8\pi }\int d\tau \left ( 4\pi \exp \left( -e^{2\mathcal{X}
+\tau -\left( z+\ln 2\right) }\right) \right) e^{-e^{z}\cosh {\tau }}
\exp (i\tau \mathcal{P})  \nonumber \\   
&=&\frac{1}{2}\int d\tau \,\exp \left( -\frac{1}{2}e^{2\mathcal{X}+\tau -z}-
\frac{1}{2}e^{z+\tau }-\frac{1}{2}e^{z-\tau }\right) 
\exp (i\tau \mathcal{P})   ~.    
\ee   
This is the operator correspondent to (\ref{StarG}); it reflects the 
Weyl correspondence through which it was originally defined (although,
technically, it was taken out of Weyl ordering above, merely as a 
matter of convenience, not a bona-fide change of representation). 

This form  leads to a more intuitive Hilbert space representation. 
Acting to the right of a position eigen-bra, 
$\left\langle x\right| \mathcal{X}=\left\langle x\right| x$, while the 
subsequent exponential of the momentum operator just translates, 
$\left\langle x\right| \exp (i\tau \mathcal{P})=\left\langle x+\tau \right|$. 
 So the full right-operation of $\mathfrak{G}$ is 
\be 
\left\langle x\right| \mathfrak{G}(z;\mathcal{X},\mathcal{P}) &=&
\frac{1}{2}\int  d\tau \,\left\langle x+\tau \right| 
\exp \left( -\frac{1}{2}e^{2x+\tau -z}-
\frac{1}{2}e^{z+\tau }-\frac{1}{2}e^{z-\tau }\right) \nonumber  \\ 
&=&\frac{1}{2}\int dy\,\left\langle y\right| \exp \left( -\frac{1}{2} 
e^{x+y-z}-\frac{1}{2}e^{z+y-x}-\frac{1}{2}e^{z-y+x}\right). 
\ee 

Inserting $1=\int dx\,\left| x\right\rangle \left\langle x\right| $  
gives $\mathfrak{G}(z;\mathcal{X},\mathcal{P})=\int dx\,\left| x\right
\rangle  \left\langle x\right| \mathfrak{G}(z;\mathcal{X},\mathcal{P})$, 
and leads to a coordinate space realization of the operator 
involving an $x,y$-symmetric kernel, 
\be 
\mathfrak{G}(z;\mathcal{X},\mathcal{P})=\frac{1}{2}\int dxdy\,\left|  
x\right\rangle \left\langle y\right| \,\exp \left( -\frac{1}{2}e^{x+y-z}-
\frac{1}{2}e^{x-y+z}-\frac{1}{2}e^{-x+y+z}\right)  . 
\label{CoordSpaceGOperator}
\ee 

This operator is diagonal on energy states: by Macdonald's identity 
(\ref{TrilinearMacdonald}), and the reality and orthogonality of the wave 
functions, 
\be 
\left\langle E_{1}\right| \mathfrak{G}(z;\mathcal{X},\mathcal{P})\left| 
E_{2}\right\rangle  &=&\frac{1}{2}\int dxdy\,\psi _{E_{1}}^{\ast }\left( 
x\right) \,\psi _{E_{2}}\left( y\right) \,\exp \left( -\frac{1}{2}e^{x+y-z}-
\frac{1}{2}e^{z+y-x}-\frac{1}{2}e^{z-y+x}\right) \nonumber  \\  
&=&\delta \left( E_{1}-E_{2}\right) \,K_{i\sqrt{E_{1}}}(e^{z})\;. 
\ee 
This is in agreement with the corresponding phase-space expression, 
(\ref{StarK}).  

The composition law of this operator also parallels its phase-space isomorph, 
(\ref{calcomp}),  
\be
\mathfrak{G}(u) \mathfrak{G}(v) ={\frac{1}{2}}\int \!dw~\exp 
\left( -{\frac{1}{2}}\left( e^{u+v-w}+e^{u-v+w}+e^{-u+v+w}\right) \right) ~
\mathfrak{G}(w).
\ee

\end{document}